\documentclass[superscriptaddress,aps,prl,twocolumn]{revtex4}
\usepackage{bm}
\usepackage{ulem}
\usepackage{epsfig}
\usepackage{graphicx}
\usepackage{amssymb,amsmath,amsbsy,amsgen,amsfonts}    
\usepackage{dcolumn}
\usepackage{amsthm}
\usepackage{mathrsfs}
\usepackage{latexsym}
\usepackage{array}
\usepackage{color}
\usepackage{amstext}
\allowdisplaybreaks[1]
\usepackage{txfonts}
\usepackage{pifont}

\usepackage{epstopdf} 

\newcommand{\bra}[1]{\left\langle{#1}\right\vert}
\newcommand{\ket}[1]{\left\vert{#1}\right\rangle}

\newcommand{\be}{\begin{equation}}
\newcommand{\ee}{\end{equation}}
\newcommand{\ba}{\begin{array}}
\newcommand{\ea}{\end{array}}
\newcommand{\bqa}{\begin{eqnarray}}
\newcommand{\eqa}{\end{eqnarray}}

\DeclareSymbolFont{symbols}{OMS}{cmsy}{m}{n}

\begin{document}

\title{Probing decoherence in plasmonic waveguides in the quantum regime}

\author{S. G. Dlamini}
\thanks{These authors contributed equally}
\affiliation{School of Chemistry and Physics, University of KwaZulu-Natal, Durban 4001, South Africa}
\author{J. T. Francis}
\thanks{These authors contributed equally}
\affiliation{School of Chemistry and Physics, University of KwaZulu-Natal, Durban 4001, South Africa}
\author{X. Zhang}
\affiliation{School of Chemistry and Physics, University of KwaZulu-Natal, Durban 4001, South Africa}
\author{\c{S}. K. \"Ozdemir}
\affiliation{Department of Engineering Science and Mechanics, Pennsylvania State University, University Park, Pennsylvania 16802, USA}
\author{S. Nic Chormaic}
\affiliation{School of Chemistry and Physics, University of KwaZulu-Natal, Durban 4001, South Africa}
\affiliation{Light-Matter Interactions Unit, Okinawa Institute of Science and Technology Graduate University, Onna, Okinawa 904-0495, Japan}
\author{F. Petruccione}
\affiliation{School of Chemistry and Physics, University of KwaZulu-Natal, Durban 4001, South Africa}
\affiliation{National Institute for Theoretical Physics, KwaZulu-Natal, South Africa}
\author{M. S. Tame}
\email{markstame@gmail.com}
\affiliation{School of Chemistry and Physics, University of KwaZulu-Natal, Durban 4001, South Africa}

\date{\today}

\begin{abstract}
We experimentally investigate the decoherence of single surface plasmon polaritons in metal stripe waveguides. In our study we use a Mach-Zehnder configuration previously considered for measuring decoherence in atomic, electronic and photonic systems. By placing waveguides of  different length in one arm we are able to measure the amplitude damping time $T_1=1.90 \pm 0.01 \times 10^{-14}$~s, pure phase damping time $T_2^*=11.19 \pm 4.89 \times 10^{-14}$~s and total phase damping time $T_2=2.83 \pm 0.32 \times 10^{-14}$~s. We find that decoherence is mainly due to amplitude damping, and thus loss arising from inelastic electron and photon scattering plays the most important role in the decoherence of plasmonic waveguides in the quantum regime. However, pure phase damping is not completely negligible. The results will be useful in the design of plasmonic waveguide systems for carrying out phase-sensitive quantum applications, such as quantum sensing. The probing techniques developed may also be applied to other plasmonic nanostructures, such as those used as nanoantennas, as unit cells in metamaterials and as nanotraps for cold atoms.
\end{abstract}


\maketitle

\section{Introduction} 
Plasmonic systems involve electromagnetic excitations of light coupled to electron charge density oscillations on the surface of metals~\cite{Maier07}. These hybrid excitations of light and matter are known as surface plasmon polaritons (SPPs) and the electromagnetic field is highly confined~\cite{Takahara97,Takahara09}. This confinement has opened up many applications for controlling light at the nanoscale, including nanoantennas for sending and receiving light signals~\cite{Gian11}, the enhancement of photovoltaics for solar cell technology~\cite{Atwater10}, and many more~\cite{Gramotnev10}. The hybrid nature of SPPs has also raised the interesting prospect of integrating photonics and electronics in the same platform~\cite{Ozbay06}. Most recently, studies have investigated plasmonics in the quantum regime~\cite{Tame13}, with single-photon sources~\cite{Akimov07,Kolesov09,Huck11,Cuche11} and single-photon switches~\cite{Chang07,Kolchin11,Chang14} being proposed and experimentally realized. These nanophotonic devices are important for emerging quantum technologies, such as photonic-based quantum computers~\cite{OBrien05,Ladd10} and quantum communication networks~\cite{Gis02}. Following on from early work probing SPPs with quantum states of light, such as entangled photons~\cite{Alte02}, recent studies have demonstrated several key quantum applications, including quantum sensing and imaging~\cite{Fan15,Pooser15,Lee16,Lee17,Holt16}, quantum spectroscopy~\cite{Kal14}, quantum logic gates~\cite{Wang16}, entanglement generation~\cite{Dieleman17} and distillation~\cite{Asano15}, and quantum random number generation~\cite{Francis16}. What is surprising is that all of these applications can be realized even in the presence of loss, which is always present in plasmonic systems as they are scaled down to confine light to smaller scales. 

In the classical regime, loss has been studied extensively, both in plasmonic nanostructures and waveguides~\cite{Maier07}. At the microscopic level, loss is mainly due to the electron dynamics in the metal, which are governed by electron-electron scattering events, and electrons scattering with other charge carriers, phonons, defects and impurities~\cite{Link00}. In the quantum regime, loss  -- commonly referred to as amplitude damping~\cite{Nielsen00} -- has recently been studied in terms of its impact on the quantum statistics of single SPPs in waveguides~\cite{DiMartino12,Tame08}. However, in addition to loss of amplitude, an important factor that also needs to be taken into account is loss of coherence, both spatial and temporal~\cite{Mandel95}. In the classical regime, there have been many works that have investigated loss of coherence in plasmonic nanostructures and waveguides, both spatially~\cite{Schouten05,Zia07,Gan07,Morill16} and temporally~\cite{Sonn02,Kim03,Anderson10,Bosman13,Zhao16,Wang14}. At the microscopic level, pure loss of coherence is due to elastic electron scattering processes that do not lead to the loss of energy from the plasmon oscillation~\cite{Heilweil85,Sonn02}. In the quantum regime, loss of coherence -- commonly referred to as phase damping~\cite{Nielsen00} -- has not yet been studied for single SPPs. While results in the classical regime suggest that phase damping does not have a significant impact on the plasmon dynamics in nanostructures~\cite{Sonn02} and in waveguides of short length~\cite{Wang14}, it is not yet known how low-level excitations of light are affected, nor what role it may play in the plasmon dynamics in longer waveguides. Given the increasing number of applications already demonstrated for plasmonics in the quantum regime it is important to understand the relative impact of amplitude damping, which also causes loss of coherence, and phase damping, so that phase-sensitive quantum applications may be properly developed.
\begin{figure*}[t]
\includegraphics[width=17.5cm]{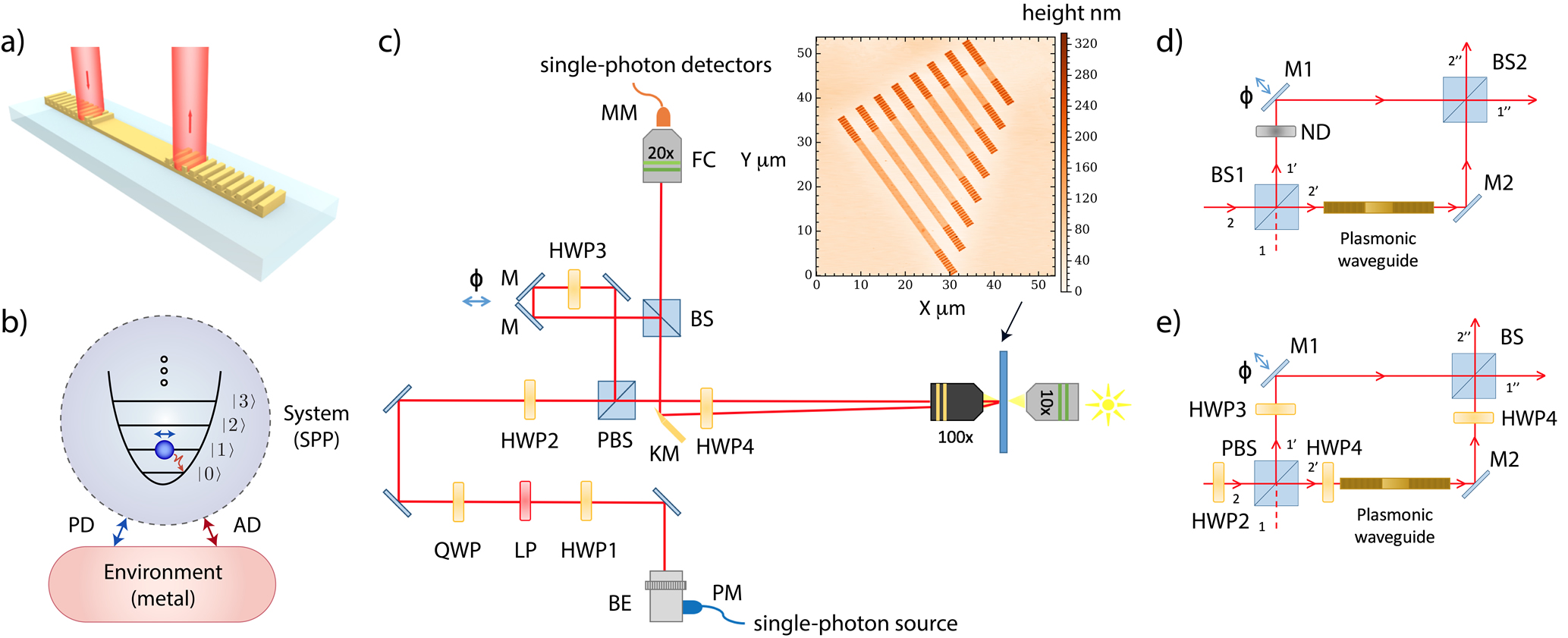}
\caption{Experimental setup for probing the decoherence of single surface plasmon polaritons (SPPs). {\bf (a)} Pictorial representation of the type of plasmonic waveguide probed. An input grating is used to couple single photons into the plasmonic waveguide, creating single SPPs which propagate along the waveguide, and then decouple back into single photons at an output grating. {\bf (b)} Diagram showing two main damping channels for the waveguides -- amplitude damping (AD) and phase damping (PD) -- and their effect on the internal number state of the bosonic SPP: AD causes a loss of energy and reduces coherence (red arrows), while PD maintains energy but reduces coherence (blue arrows). {\bf (c)} Microscope stage for probing the waveguides in configuration B, which is used for measuring phase damping. Configuration A, which is used for measuring amplitude damping, does not include half wave-plate 2 (HWP2), the polarising beamsplitter (PBS) and the 50:50 beamsplitter (BS) - see main text for details. The inset shows a three-dimensional atomic force microscope image of the different length gold stripe waveguides used. {\bf (d)} A Mach-Zehnder interferometer (MZI) for probing phase damping. {\bf (e)} A modified version of the MZI with a polarizing beamsplitter, as used in the microscope stage.}
\label{fig1} 
\end{figure*}

In this work we experimentally investigate amplitude and phase damping for single SPPs in waveguides. We refer to both types of damping as `decoherence' because amplitude damping also reduces the coherence properties of single excitations~\cite{Nielsen00,Zurek03}. For the dimensions of the gold stripe waveguides we use, as depicted in Fig.~\ref{fig1}~(a), the spatial mode is well defined as a single mode~\cite{Lamprecht01,Zia05,Zenin16,Liu16}, with the SPPs excited in the number state degree of freedom. As a result, the decoherence is in the temporal domain as the SPP propagates. We probe plasmonic waveguides of varying lengths in a Mach-Zehnder interferometer configuration that has previously been used to study decoherence in atomic~\cite{Chapman95,Koko01,Bertet01,Uys05,Cronin09}, electronic~\cite{Roulleau08,Neuenhahn08,Haack11}, photonic~\cite{Schwindt99,Jacques08} and relativistic~\cite{Zych11} quantum systems. The configuration allows us to extract out values for the two main damping mechanisms of the SPP system, as depicted in Fig.~\ref{fig1}~(b): the amplitude damping time, $T_1$ -- the time it takes for the probability of an SPP in the excited state to reduce to $1/e$ its initial value -- and the pure phase damping time, $T_2^*$ -- the time it takes for the off-diagonal elements of an SPP state to reduce to $1/e$ their initial values. The total phase damping time, $T_2$, for a single SPP includes contributions from both $T_1$ and $T_2^*$, and is given by the relation $T_2^{-1}=T_1^{-1}/2+T_2^{*~-1}$, {\it i.e.} $T_2 \leq 2T_1$~\cite{Nielsen00}, where the presence of $T_1$ is a result of amplitude damping also contributing to total phase damping. In our experiment we find values of $T_1=1.90 \pm 0.01 \times 10^{-14}$~s, $T_2^*=11.19 \pm 4.89 \times 10^{-14}$~s and therefore $T_2=2.83 \pm 0.32 \times 10^{-14}$~s. These suggest that the total phase damping time is dominated by amplitude damping, showing that loss of amplitude is the most important factor in the decoherence of single SPPs in the plasmonic waveguides. However, the role of pure phase damping is not completely negligible. Our work shows that both amplitude and pure phase damping can lead to decoherence in quantum plasmonic systems, and it provides useful information about the loss of coherence that should be considered when designing plasmonic waveguide systems for phase-sensitive quantum applications, such as quantum sensing~\cite{Fan15,Pooser15,Lee16,Lee17} and quantum imaging~\cite{Lee17,Holt16}. The techniques developed here for characterising decoherence in plasmonic waveguides may be useful for studying other plasmonic nanostructures, such as those used as nanoantennas~\cite{Gian11}, as unit cells in metamaterials~\cite{Soukoulis11,Meinzer14} and as nanotraps for cold atoms~\cite{Stehle14}.

\section{Experimental setup} 
The setup used to probe SPP decoherence is shown in Fig.~\ref{fig1}c. Here, a microscope is used to excite single SPPs on plasmonic waveguides by coupling in single photons generated via spontaneous parametric down-conversion (SPDC). Pairs of horizontally polarized single photons at $810$~nm are produced by using a vertically polarized $200$~mW continuous wave laser at $405$~nm focused onto a Beta Barium Borate (BBO) crystal cut for type-I SPDC. Phase matching conditions lead to photons from a given pair being emitted into antipodal points of a forward directed cone with an opening angle of $6^\circ$~\cite{Burnham70,Hong86}. Polarizing beamsplitters (PBSs) are positioned in the path of the down-converted beams to clean up the polarization of the photons and remove any light with vertical polarization. Filters at $800$~nm are placed on both paths ($\Delta \lambda=40$~nm) to spectrally select out the down-converted photons. Such broad filters are used in order to maximize the generation rate of photon pairs for probing the plasmonic waveguides. While this influences the spectral quality of the photons, we will show later that we obtain a second-order correlation value well below 0.5, which is a clear indication that our experiments are performed in the single-photon regime. After the filters, each beam from the SPDC is sent to a single-mode fiber. One of the fibers is directly connected to a single-photon silicon avalanche photodiode detector (SAPD) Excelitas SPCM-AQR-15, which monitors the arrival of one photon from a given SPDC pair. A detection of a photon at the SAPD heralds the presence of a single photon in the other fiber~\cite{Hong86}. In order to maintain the polarization of the heralded photon while it is transferred to the microscope, a polarization maintaining (PM) fiber is used. 

Two main configurations of the setup shown in Fig.~\ref{fig1}c are used in the experiment. We denote these as configuration A and configuration B. In configuration A, which is used for measuring amplitude damping, half-wave plate 2 (HWP2), the PBS and the beamsplitter (BS) are not present. In this case, single photons are introduced to the stage via the beam expander (BE). Then, HWP1, a linear polarizer (LP) and a quarter-wave plate (QWP) are used to control the polarization of the photons and maintain them as linearly polarized. HWP4 is used to optimize the polarization for coupling the single photons into single SPPs on the waveguides~\cite{DiMartino12}. A microscope objective (100$\times$) focuses the beam of single photons onto the input grating of a plasmonic waveguide, as depicted in Fig.~\ref{fig1}a. Excited single SPPs then propagate along the waveguide and are decoupled back into photons at an output grating. The microscope collects the decoupled photons, which are picked off by a knife-edge mirror (KM) and directed to a multimode fiber (MM) via a fiber coupler (FC). The MM fiber is connected to a SAPD. A detection of a photon together with a detection of the corresponding heralding photon from the SPDC pair within a coincidence window of $8$~ns confirms single photons were sent through the microscope stage, converted to SPPs and then back into photons again. 

In configuration B, which is used for measuring phase damping, all components shown in Fig.~\ref{fig1}c are present. These enable the quantification of the impact of waveguide propagation on the coherence properties of single photons converted into SPPs. In this configuration, the microscope becomes part of one arm in a Mach-Zehnder interferometer (MZI) by using the PBS and BS, with one path photonic and the other plasmonic. Details of configuration B will be described later.

The plasmonic waveguides probed have a range of different lengths, from $7.32~\mu$m to $32.47~\mu$m. They are gold stripes $2~\mu$m wide and 70~nm high. At the ends of the waveguides are gratings of height 90~nm made from 11 steps of period 740~nm, serving as inputs and outputs for converting photons to SPPs and back again~\cite{DiMartino12}. Due to the design of the gratings, the optimal angle for in-coupling a photon is normal to the waveguide surface. Furthermore, due to reciprocity, the photons output from a grating at the end of a waveguide are also normal to the waveguide. This enables the insertion and collection optics in our setup to all be placed on the same side of the waveguide sample. The waveguides are fabricated as follows. First, a positive photoresist is spin-coated on a silica glass substrate (refractive index 1.526), and then electron beam lithography is used to define the waveguide regions. Finally, a lift-off technique is used, with an adhesion layer of Ti (thickness 2-3 nm) followed by a 70~nm Au layer deposition using electron beam evaporation. The gratings are formed on the top in a similar process, utilising alignment marks to match the layers. A 3D image of the waveguides has been obtained using an atomic force microscope (NT-MDT Smena), as shown in the inset of Fig.~\ref{fig1}c.

\section{Results}
We start with the results for amplitude damping of single SPPs using the microscope stage in configuration A, {\it i.e.}~without the MZI (HWP2, PBS and BS removed). Recent experiments have confirmed the bosonic nature of SPPs~\cite{Heeres13, Fakonas14, DiMartino14, Cai14, Fujii14} and explored related quantum behaviour~\cite{Vest17}. Initial results have also been obtained for amplitude damping of single SPPs~\cite{DiMartino12}. Here, we confirm these results and provide a more detailed analysis of the role of amplitude damping in the decoherence process. We then investigate phase damping of single SPPs, which to our knowledge has not been done before. The study of amplitude and phase damping at the same time allows us to combine both into a general model for decoherence of single SPPs. In Fig.~\ref{fig1}b we show the energy level structure for a system of a bosonic particle (the SPP)~\cite{Loudon00}. Amplitude damping is associated with energy loss and the system, initially in an excited state $\ket{1}$, will decay to the ground state $\ket{0}$ after some time $t$ through its interaction with the environment. For the SPP this arises from electron collisions in the supporting metal which cause energy loss in the electronic degree of freedom of the SPP, as well as surface defects and the mode structure of the waveguide causing energy loss in the optical degree of freedom due to coupling of light into the far-field. In general, for single bosonic excitations undergoing amplitude damping we have the following transformation of the density matrix for the system,
\be
\rho(0) \to \rho(t)= \left( \begin{array}{cc} \rho_{00} +(1-e^{-\Gamma_1 t})\rho_{11} & e^{-\Gamma_1 t/2}\rho_{01} \\ e^{-\Gamma_1 t/2} \rho_{10} & e^{-\Gamma_1 t}\rho_{11} \\ \end{array} \right),
\label{AD}
\ee
where $\rho_{ij}=\bra{i} \rho(0) \ket{j}$ are the initial entries of the density matrix at $t=0$ in the number state basis, $\ket{n}$, and $\Gamma_1$ characterizes the strength of the damping induced by the environment~\cite{Nielsen00}. In the classical regime, $\Gamma_1$ corresponds to population decay or loss, the value of which is easily found by measuring the decay of the SPP intensity as a function of waveguide length. Here, the length at which the intensity has dropped to $1/e$ of its initial value is the propagation length $L$~\cite{Maier07}, and the value for $\Gamma_1$ is then the inverse of the time at which the SPP reaches this length ($T_1$), given by $\Gamma_1=v_g/L$, where $v_g$ is the group velocity of the SPP. In the quantum regime, when single SPPs are considered, the value of $\Gamma_1$ can be found similarly, but the intensity measurement is replaced by the mean single-excitation count rate~\cite{DiMartino12,Tame08}. This can be obtained in our setup by measuring the rate of coincidences between the heralding photon and the photon that has undergone the photon-SPP-photon conversion process, as the waveguide length increases. A coincidence detection corresponds to the case where a single photon was generated, converted to a single SPP and then converted back to a single photon. The length at which the coincidence rate drops to $1/e$ of its initial value is then the propagation length $L$ in the single-SPP regime. It represents the length at which the probability of an excited single SPP to propagate to that point reduces to $1/e$~\cite{DiMartino12}. The value for $\Gamma_1$ is then obtained as in the classical case. To check that we are able to probe single SPPs in the waveguides we measure the second-order correlation function $g^{2}(0)$ for single photons sent through a waveguide of length $7.47~\mu$m, as described in Ref.~\cite{DiMartino12}. We find $g^{(2)}(0)= 0.26\pm 0.01$, which is below $0.5$, confirming we are in the single-excitation regime~\cite{Loudon00}. 
\begin{figure}[t]
\includegraphics[width=8.7cm]{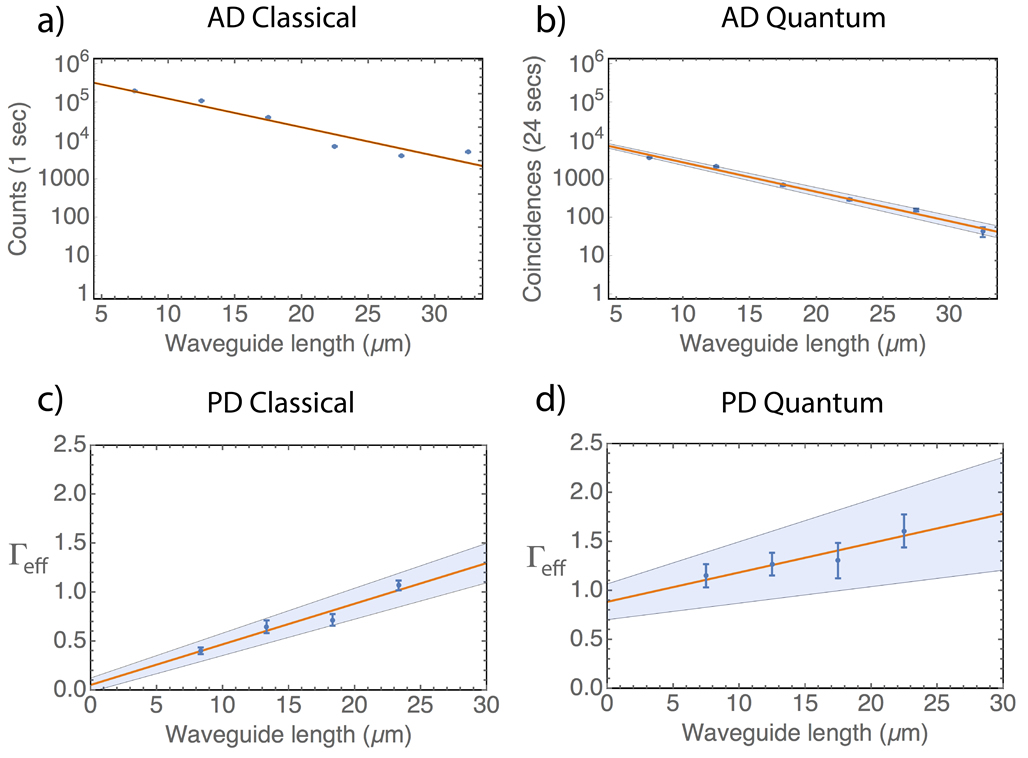}
\caption{Decoherence in the classical and quantum regime. {\bf (a)} Intensity throughput as a function of waveguide length showing amplitude damping for classical SPPs. {\bf (b)} Amplitude damping for single SPPs in the quantum regime measured via coincidences with a heralding photon. {\bf (c)} Effective phase damping parameter $\Gamma_{eff}$ as a function of waveguide length showing pure phase damping for classical SPPs. {\bf (d)} Effective phase damping parameter $\Gamma_{eff}$ showing pure phase damping for single SPPs. The shaded regions represent upper and lower values of a straight line best fit using the least squares method and a Monte Carlo simulation drawing each data point from within its individual standard deviation with Poissonian distribution.}
\label{fig2} 
\end{figure}

We first measure the propagation length $L$ using the microscope in configuration A in the classical regime using a white laser source (Fianium WL-MICRO) and a filter centred at $810$~nm with $\Delta \lambda=10$~nm. The input intensity is set to a few mW and the transmitted light intensity ($10^{4} - 10^{5}~{\rm cps}$) is recorded by an SAPD coupled to a MM fiber, as shown in Fig.~\ref{fig1}~(c). The results for different waveguide lengths are shown in Fig.~\ref{fig2}~(a). One can clearly see the well-observed exponential decay of the intensity as the waveguide length increases. We find a propagation length of $L=5.85\pm 0.03~\mu$m. This value is similar to previous experimental work~\cite{DiMartino12,Francis16}, although slightly smaller than the $10~\mu$m predicted using finite element simulation (COMSOL) of the stripe waveguide~\cite{Lamprecht01,Zia05,Zenin16,Liu16}. The difference may be caused by edge effects along the lateral width of the waveguides, surface and material defects during fabrication, and a small deviation of the actual dielectric function of gold from that used in the simulation~\cite{Rakic98}. 

To convert the propagation length to the amplitude damping time $T_1$ we obtain the SPP dispersion relation for the plasmon mode in the waveguide from the simulation. Based on this, we find the group velocity $v_g(\omega_0)=2.958 \times 10^{8}~{\rm ms}^{-1}$ at the free-space wavelength $\lambda_0=810$~nm. A more rigorous approach would be to directly measure the group velocity; however, for the waveguide dimensions and free-space wavelength we consider, theoretical simulation describes the experimental data reasonably well~\cite{Zenin16}. Furthermore, here we use the group velocity simply to convert damping factors into the time domain and their values in the spatial domain and relative ratios are valid regardless. Using the group velocity we find $\Gamma_1=5.06 \pm 0.01 \times 10^{13}~{\rm s}^{-1}$ and an amplitude damping time of $T_1=\Gamma_1^{-1}=1.98 \pm 0.01  \times 10^{-14}$~s.

In Fig.~\ref{fig2}~(b) we show the results for single SPPs in our experiment. Here, the exponential decay of the mean count rate (observed via the coincidence rate) is seen as the waveguide length increases. The data collection time has been increased to 24s for each length in order to measure a similar number of counts as the classical case, which has a shorter collection time of 1s. We find a propagation length of $L=5.61\pm 0.05~\mu$m, consistent with the result from the classical regime. From this we obtain $\Gamma_1=5.27 \pm 0.02 \times 10^{13}~{\rm s}^{-1}$ and a single-SPP amplitude damping time of $T_1=\Gamma_1^{-1}=1.90 \pm 0.01 \times 10^{-14}$~s. In general, the relation between the phase damping time $T_2$ and amplitude damping time $T_1$ is given by $T_2^{-1}=T_1^{-1}/2+T_2^{*~-1}$~\cite{Nielsen00}, where $T_2^*$ is the pure phase damping time. Thus, from the above result we already have an upper bound of $T_2 \leq 2T_1$ for single SPPs in the quantum regime. However, $T_2^*$ remains to be found to determine the exact value of $T_2$, and could reduce it appreciably.

Pure phase damping characterized by the time $T_2^*$ is associated with interactions where energy is maintained and therefore a system initially in a ground state, or excited state, will remain in that state after some time $t$. However, a state in a superposition of ground and excited states will experience a loss of coherence between the states due to a time varying change in the relative phase. For the SPP this arises from electron collisions in the supporting metal associated with elastic processes~\cite{Heilweil85,Sonn02}. For single bosonic excitations we have the following transformation of the density matrix,
\be
\rho(0) \to \rho(t)= \left( \begin{array}{cc} \rho_{00} & e^{-\Gamma_2^* t}\rho_{01} \\ e^{-\Gamma_2^* t} \rho_{10} & \rho_{11} \\ \end{array} \right),
\label{PD}
\ee
where $\Gamma_2^*$ characterizes the strength of the damping induced by the environment~\cite{Nielsen00}. In the classical regime, $\Gamma_2^*$ corresponds to the loss of temporal coherence. We obtain its value in the classical and quantum regime by placing different length plasmonic waveguides inside a MZI and measuring the loss of interference between the two paths, as shown in Fig.~\ref{fig1}~(d). In what follows, we describe how this is done in the quantum regime and link it with the classical case in the corresponding limit. 

We start with the case of no decoherence in the waveguides. In Fig.~\ref{fig1}~(d) we consider the input state $\ket{0}_{1}\ket{1}_{2}$, corresponding to a single photon in mode 2. The first beam splitter (BS1) transforms the state to~\cite{Loudon00}
\be
\frac{1}{\sqrt{2}}(\ket{0}_{1'}\ket{1}_{2'}+i\ket{1}_{1'}\ket{0}_{2'}). \label{BS1}
\ee
Taking the neutral density (ND) filter and plasmonic waveguide as having unit transmission, and the mirrors (M1 and M2) contributing a phase factor $e^{i \pi/2}$ to each term, we have the following state after the second beamsplitter (BS2),
\be
\frac{1}{2}[(1-e^{i(\phi-\delta)})\ket{0}_{1''}\ket{1}_{2''}+i(1+e^{i(\phi-\delta)})\ket{1}_{1''}\ket{0}_{2''}].
\ee
Here, the phase $\phi$ corresponds to a change in path length $1'$ caused by mirror M1 placed on a translation stage and the phase $\delta=k_{spp}\ell$, with $k_{spp}$ the SPP wavenumber and $\ell$ the length of the plasmonic waveguide. The probability of a photon detected in mode $1''$ is then simply
\begin{equation}
p(\phi)=\frac{1}{2}(1+\cos(\phi-\delta)).
\end{equation}

We now introduce decoherence in the system. When amplitude and pure phase damping are included in the plasmonic waveguide, the transformations in Eqs.~(\ref{AD}) and (\ref{PD}) are applied to the state after the first beamsplitter, given by Eq.~(\ref{BS1}). The transformations are given explicitly for mode $2'$ by $\ket{0}\bra{0} \to \ket{0}\bra{0}+(1-e^{-\Gamma_1 t})\ket{1}\bra{1}$, $\ket{0}\bra{1} \to e^{-\Gamma_2^* t}e^{-\Gamma_1 t/2}\ket{0}\bra{1}$, $\ket{1}\bra{0} \to e^{-\Gamma_2^* t}e^{-\Gamma_1 t/2}\ket{1}\bra{0}$ and $\ket{1}\bra{1} \to e^{-\Gamma_1 t}\ket{1}\bra{1}$. The probability of a photon detected in mode $1''$ then becomes
\begin{equation}
p(\phi)=\frac{1}{4}(1+e^{-\tilde{\Gamma}_1}+2e^{-\tilde{\Gamma}_1/2-\tilde{\Gamma}_2^*}\cos(\phi-\delta))
\end{equation}
where $\tilde{\Gamma}_1=\Gamma_1\ell /v_g=\ell/L$ and $\tilde{\Gamma}_2^*=\Gamma_2^*\ell/v_g$. As $\tilde{\Gamma}_1$ is already known from previous measurements and $\delta$ is a fixed phase for a given waveguide length $\ell$, then by measuring $p(\phi)$ as $\phi$ is varied using the translation stage of M1, the remaining unknown parameter $\tilde{\Gamma}_2^*$ can be extracted to obtain $\Gamma_2^*$, and thus $T_2^*$. In practice, however, the impact of amplitude damping in the plasmonic waveguide reduces the average value of $p(\phi)$ significantly and in the most extreme case we have $ p(\phi)=1/4$, as only photons going through the free-space arm of the MZI will be detected. As the amplitude damping in the plasmonic waveguide becomes large it is difficult to observe oscillations in $p(\phi)$ and extract out $\tilde{\Gamma}_2^*$. This problem can be addressed by introducing an additional tuneable amplitude damping on the free-space arm using a variable neutral density (ND) filter. As the photon is also a boson, we can use Eq.~(\ref{AD}) to model the damping, which changes the probability of detection to 
\begin{equation}
p(\phi)=\frac{1}{4}(e^{-\Gamma}+e^{-\tilde{\Gamma}_1}+2e^{-(\Gamma+\tilde{\Gamma}_1)/2-\tilde{\Gamma}_2^*}\cos(\phi-\delta)),
\end{equation}
where $\Gamma$ characterizes the amplitude damping on the free-space arm. This parameter can be tuned to match $\tilde{\Gamma}_1$ in the plasmonic waveguide by blocking the plasmonic waveguide arm and measuring the output counts in mode $1''$ as the ND filter is varied.

In order to integrate the MZI of Fig.~\ref{fig1}~(d) into our microscope stage more easily we replace BS1 and the variable ND filter with a PBS preceded by HWP2, as shown in Fig.~\ref{fig1}~(e). This configuration provides polarization control over the relative splitting into modes $1'$ and $2'$, and allows us to increase the rate of photons injected into the plasmonic waveguide compared to the original configuration of Fig.~\ref{fig1}~(d). HWP4 provides polarization control for optimising coupling of single photons to single SPPs and HWP3 rotates the polarization of the free-space arm to match the output of the plasmonic beamsplitter in order to obtain interference at the BS. For a given waveguide length, once HWP3 and HWP4 have been modified, the polarization state in the free-space and plasmonic arms is fixed for the entire set of measurements. The above modifications change the detection probability to
\begin{equation}
p(\phi)=\frac{1}{2}(e^{-\Gamma_{1'}}+e^{-(\tilde{\Gamma}_1+\Gamma_{2'})}+2e^{-(\Gamma_{1'}+\Gamma_{2'}+\tilde{\Gamma}_1)/2-\tilde{\Gamma}_2^*}\cos(\phi-\delta)),
\end{equation}
where $\Gamma_{1'}$ and $\Gamma_{2'}$ are controlled by HWP2, and we set $\Gamma_{1'}=\tilde{\Gamma}_1+\Gamma_{2'}$ in order to observe clearly a symmetric oscillation in $p(\phi)$. Finally, we include a possible asymmetry in the splitting at the BS, which has an order of magnitude larger error in its splitting than the PBS. With reflection and transmission coefficients $R$ and $T$, respectively, for the BS, this changes the detection probability to
\begin{eqnarray}
p(\phi)&=&R \, e^{-\Gamma_{1'}}+T \, e^{-(\tilde{\Gamma}_1+\Gamma_{2'})} \nonumber \\
&&+2\sqrt{RT}e^{-(\Gamma_{1'}+\Gamma_{2'}+\tilde{\Gamma}_1)/2-\tilde{\Gamma}_2^*}\cos(\phi-\delta).
\label{pphi}
\end{eqnarray}
From the above equation it would appear that only a single waveguide length is needed to extract out $\tilde{\Gamma}_2^*$. However, in practice it is not always possible to get a complete overlap of modes $1'$ and $2'$ at the BS. This non-ideal overlap reduces the visibility of the oscillations in $p(\phi)$ and acts as an effective phase damping, which we describe using the parameter $\Gamma_{int}$. Thus, $\tilde{\Gamma}_2^*$ in Eq.~(\ref{pphi}) is transformed as $\tilde{\Gamma}_2^* \to \Gamma_{eff}=\tilde{\Gamma}_2^*+\Gamma_{int}$. Due to this non-ideal overlap, it appears that we must also find $\Gamma_{int}$ to obtain $\tilde{\Gamma}_2^*$. This can be done by extracting $\Gamma_{eff}$ from $p(\phi)$ for waveguides of different lengths and then using $\Gamma_{eff}(\ell)=\Gamma_2^*\ell/v_g+\Gamma_{int}$, where the pure phase damping per unit length, $\Gamma_2^*/v_g$, is the gradient of $\Gamma_{eff}(\ell)$ and $\Gamma_{int}$ is the $y$-intercept. 

\begin{figure}[t]
\includegraphics[width=9.1cm]{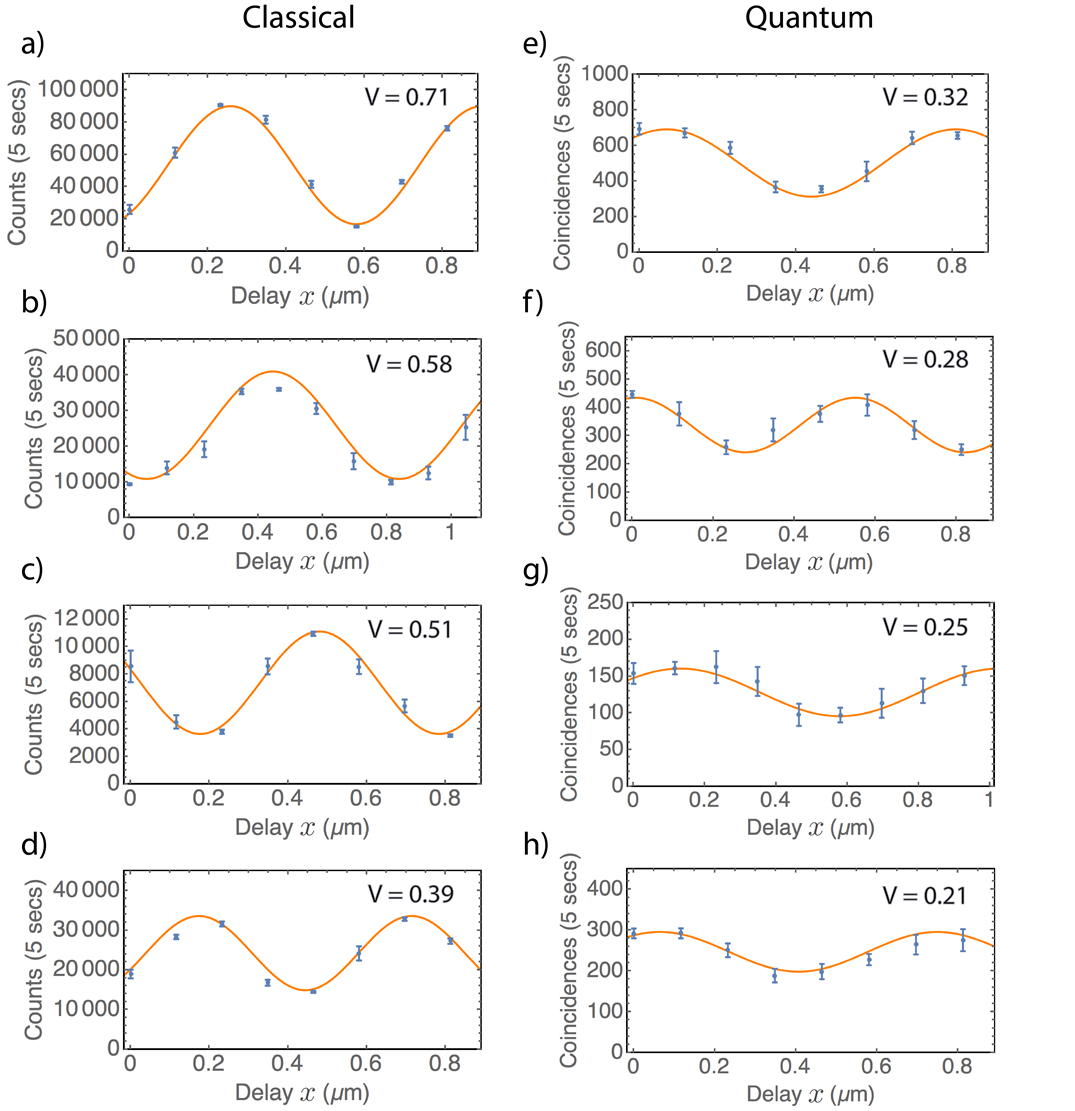}
\caption{Intensity dependence in the classical regime and coincidence counts in the quantum regime for the output signal from the MZI for different waveguide length as the phase $\phi$ is modified. Here, $\phi=2 \pi s x/\lambda_0$, where $s$ accounts for the translation stage geometry and $x$ is its position $\times 2$ (total delay). {\bf (a)-(d)} The left hand column corresponds to the classical regime with intensity measured as counts. {\bf (e)-(h)} The right hand column corresponds to the quantum regime with intensity measured as coincidences. The solid lines are fits using $p(\phi)$. The length of the waveguide increases with row number in steps of 5~$\mu$m and is 8.31~$\mu$m, 13.31~$\mu$m, 18.31~$\mu$m and 23.31~$\mu$m for the left hand column and 7.47~$\mu$m, 12.47~$\mu$m, 17.47~$\mu$m and 22.47~$\mu$m for the right hand column. The visibility is given in the inset for each panel and related to system parameters by $V=(p_{max}-p_{min})/(p_{max}+p_{min})$. While $\Gamma_1$ has been obtained using a fixed input intensity for all waveguide lengths, a fixed input intensity is not used here. This is due to the additional presence of the interferometer, which leads to the oscillating output signal being more sensitive as the measured counts reduce. As a result, the input intensity is increased for longer waveguide lengths using HWP's 3 and 4, which are fixed for the entire set of measurements for a specific waveguide. Therefore maximum counts do not necessarily decrease as the waveguide length increases. $I_{in}$ is modified in the extraction model to take this into account.}
\label{fig3}
\end{figure}

In Fig.~\ref{fig2}~(c) and (d) we plot $\Gamma_{eff}(\ell)$ for increasing waveguide length in the classical and quantum regime, respectively. For the classical case, $\Gamma_{eff}(\ell)$ is obtained by fitting the function $I(\phi)=I_{in} p(\phi)$ to intensity measurements, where $I_{in}$ is the initial input intensity to the MZI. Examples of the intensity measurements for the different waveguide lengths probed in the classical regime are shown in Fig.~\ref{fig3}~(a)-(d) (left hand column) over a period of oscillation. A Monte Carlo simulation is carried out for each of these figures, where $\Gamma_{eff}(\ell)$ is varied to fit the function $I(\phi)$ for 200 instances of a given figure. Each instance has its data points drawn randomly from within the standard deviations measured at each value of $\phi$ using a Poissonian distribution. All other parameters of $I(\phi)$ are known except for $\Gamma_{eff}(\ell)$, and the resulting values extracted are shown in Fig.~\ref{fig2}~(c). The error bars on each value are obtained by analysing and fitting $I(\phi)$ to several periods of oscillation for each waveguide length $\ell$. From Fig.~\ref{fig2}~(c) we find a gradient of $\Gamma_2^*/v_g=0.042 \pm 0.003~(\mu {\rm m})^{-1}$ and thus a value of $\Gamma_2^*=1.25 \pm 0.11 \times 10^{13}~{\rm s}^{-1}$ and $T_2^*=8.03 \pm 0.71 \times 10^{-14}$~s. 

It should be noted that the periods of the oscillations shown in Fig.~\ref{fig3} are not all equal to the wavelength of the single photons (810~nm). The change in the period is due to small differences in the angle of the output beam for different length waveguides. Although the output beams from the gratings are designed to be normal to the waveguide surfaces, small differences in the lateral beam displacement due to the different length of the waveguides results in an angle change when the beams pass through the microscope objective. The result is that the delay distance $x$ that the mirror stage moves is rescaled by a small geometric factor $s$, becoming $sx$. The change in period does not have any effect on the values of the decay parameters extracted from the fits as these are dependent only on the amplitude and mid-point of the oscillations.

In Fig.~\ref{fig2}~(d) we perform the same extraction method for single SPPs and Fig.~\ref{fig3}~(e)-(h) (right hand column) shows examples of the oscillations used for each waveguide length. From Fig.~\ref{fig2}~(d) we find a gradient of $\Gamma_2^*/v_g=0.030 \pm 0.013~(\mu {\rm m})^{-1}$ and thus a value of $\Gamma_2^*=0.89 \pm 0.39 \times 10^{13}\,\,{\rm s}^{-1}$ and $T_2^*=11.19 \pm 4.89 \times 10^{-14}$~s. While the results from the quantum case are clearly statistically more noisy, the values are consistent with those found in the classical regime to within a standard deviation. 

It is also interesting to inspect the values of $\Gamma_{int}$, which are found to be $0.048 \pm 0.061$ and $0.893 \pm 0.193$ for the classical and quantum case. The difference in values is due to the better mode overlap achieved in the classical case, as the interference could be optimized by monitoring the intensity fluctuations with a spectrometer in real-time and with a reduced bandwidth for the source of light. Indeed, one can see the better mode overlap via the high visibility of the oscillations in the classical case in Fig.~\ref{fig3}~(a). For the quantum case, due to the low count rates real-time monitoring could not be performed and a similarly good mode overlap was not possible. The low count rates are also the cause of the larger error bars in Fig.~\ref{fig2}~(d), as the statistical fluctuations are larger due to the instability of the MZI over the longer time periods required for data collection. The single-SPP amplitude damping measurements shown in Fig.~\ref{fig2}~(b) do not require the MZI and thus have smaller error. Improvements to the generation rate of our single-photon source would allow an increase in visibility and reduction in the error in the phase damping investigation. It would also allow the probing of longer waveguides. However, even with the current setup we are able to observe the same trend of $\Gamma_{eff}$ in the quantum regime as seen in the classical regime.
\begin{table}[b]
\centering
\begin{center}
\noindent\begin{tabular*}{\columnwidth}{@{\extracolsep{\stretch{1}}}*{4}{r}@{}}
  \hline
  & Classical~~~~~~~~~~~ & Quantum~~~~~~~~~~~ \\
  \hline
  \hline
$\Gamma_1$ & $5.06 \pm 0.01 \times 10^{13}~{\rm s}^{-1}$ & $5.27 \pm 0.02 \times 10^{13}~{\rm s}^{-1}$ \\
$\Gamma_2^*$ & $1.25 \pm 0.11 \times 10^{13}~{\rm s}^{-1}$ & $0.89 \pm 0.39 \times 10^{13}~{\rm s}^{-1}$ \\
$\Gamma_2$ & $3.77 \pm 0.12 \times 10^{13}~{\rm s}^{-1}$ & $3.53 \pm 0.40 \times 10^{13}~{\rm s}^{-1}$ \\
\hline
$T_1$ & $1.98 \pm 0.01  \times 10^{-14}$~s ~& $1.90 \pm 0.01 \times 10^{-14}$~s~ \\
$T_2^*$ & $8.03 \pm 0.71 \times 10^{-14}$~s ~& $11.19 \pm 4.89 \times 10^{-14}$~s~ \\
$T_2$ & $2.65 \pm 0.08 \times 10^{-14}$~s ~& $2.83 \pm 0.32 \times 10^{-14}$~s~ \\
\hline            
\end{tabular*}
  \end{center}    
\caption{Summary of results from probing decoherence in plasmonic waveguides.}
\label{tab:deco} 
\end{table}

An important factor that might influence our measurement of pure phase damping is dispersion in the plasmonic waveguides. For large dispersion, the SPP wavepacket would spread significantly and any interference between the photon it is converted into and the free-space photon would be reduced, and appear as phase damping. In order to assess the impact of this effect, we calculate the group velocity dispersion (GVD) coefficient, defined as $D_{\omega_0}=\frac{{\rm d}}{{\rm d} \omega}(\frac{1}{v_g(\omega)})|_{\omega_0}$~\cite{Saleh07}. Using the dispersion relation for the plasmonic waveguides from the mode simulation~\cite{Lamprecht01,Zia05,Zenin16,Liu16}, we find $D_{\omega_0}=5.81 \times 10^{-25}$~s/m-Hz. To see how this affects the interference, as an example we take an initial Gaussian wavepacket spectral amplitude for a single SPP centred on $\omega_0$ as $\xi_0(\omega)=(2 \pi \sigma_\omega^2)^{-1/4}e^{-(\omega-\omega_0)^2/4 \sigma_\omega^2}$, where a single SPP is described as $\ket{1_{\xi}}=\int {\rm d} \omega \xi_0(\omega) \hat{a}^\dag(\omega)\ket{0}$~\cite{Tame08,Loudon00}. The initial temporal spread is $\sigma_{t_0}=1/2\sigma_\omega$. After time $t$, the wavepacket has moved a distance $\ell$ and spread according to $\sigma_t=(\sigma_{t_0}^2 +(\ell D_{\omega_0}/2 \sigma_{t_0})^2)^{-1/2}$. We then have the corresponding spectral amplitude $\xi_t(\omega)=(2 \pi \sigma_{\omega,t}^2)^{-1/4}e^{-(\omega-\omega_0)^2/4 \sigma_{\omega,t}^2}$, with $\sigma_{\omega,t}=1/2 \sigma_t$. Calculating the overlap of $\xi_0(\omega)$ and $\xi_t(\omega)$ gives a quantity that represents how well the mode from the plasmonic waveguide overlaps with the free-space photonic mode at the BS in the MZI~\cite{Ozdemir02}. Here, $\xi_0(\omega)$ represents the photon in the free-space mode (negligible dispersion) and $\xi_t(\omega)$ represents the photon from the plasmonic waveguide (with dispersion). Setting $\sigma_{\omega}=\Delta_\omega/2\sqrt{2 {\rm ln} 2}$, with $\Delta_\omega$ corresponding to a FWHM of $\Delta \lambda = 40$~nm, and taking $\omega_0$ corresponding to the central wavelength $\lambda_0=810$~nm and using the GVD coefficient together with a length $\ell=90~\mu$m (more than 3 times the longest waveguide considered), we find $\int \xi_0^*(\omega)\xi_t(\omega) {\rm d} \omega=0.99$. Thus it is expected that there is a negligible impact of dispersion on the interference for the waveguide lengths considered.

We now combine all the results in this study, taking the amplitude damping and pure phase damping values found. The combined phase damping time is $T_2=(T_1^{-1}/2+T_2^{*~-1})^{-1}=2.65 \pm 0.08 \times 10^{-14}$~s and $2.83 \pm 0.32 \times 10^{-14}$~s in the classical and quantum regimes, respectively. We are therefore able to confirm that in both cases, amplitude damping is the main source of phase and amplitude decay in the plasmonic waveguides, although pure phase damping modifies the phase damping by a relatively small amount. A summary of the main results of the study is given in Tab.~\ref{tab:deco}.

While the present work has focused on a specific type of metal, {\it i.e.} gold, as an initial study, the probing of decoherence in other types of metallic media that support surface plasmons, such as silver and graphene, would be an important next step. In addition, we have considered only a fixed wavelength of $810$~nm, mainly due to experimental access to single photons at this wavelength via parametric down-conversion. However, other single-photon sources with different wavelengths are possible, such as solid state emitters, {\it e.g.} quantum dots and nitrogen vacancy centres. It is not clear what to expect at these wavelengths, as the fundamental mechanisms which cause pure phase damping in waveguides are not well known. This is an important area of future study, both theoretically and experimentally, for developing plasmonics for quantum technological applications.

\section{Discussion} 
In this work we investigated the decoherence of SPPs in plasmonic waveguides in the classical and quantum regimes. We measured both amplitude and phase damping effects of SPPs. We found that for classical SPPs and single SPPs, amplitude damping is the main source of amplitude and phase decay. The results will be useful in the design of phase-sensitive quantum plasmonic applications, such as quantum sensing and allow appropriate quantum states to be chosen for a given task to be achieved. While our work has been limited to probing decoherence for single excitations of SPPs in the quantum regime and many excitations in the classical regime, there is an intermediate regime, involving low numbers of excitations that remains to be investigated. It would be interesting to confirm the role of decoherence in this regime, where the bosonic SPP mode is treated as a qudit~\cite{Liu04}. This would be important for developing quantum plasmonic state engineering at the few SPP excitation number.

{\it Acknowledgments.---} This research was supported by the South African National Research Foundation, the National Laser Centre, the UKZN Nanotechnology Platform, the South African National Institute for Theoretical Physics, and the South African Research Chair Initiative of the Department of Science and Technology and National Research Foundation. S. K. O. acknowledges the support of the Pennsylvania State University Materials Research Institute (MRI).


\end{document}